\documentclass[review]{elsarticle}
\usepackage{float}
\usepackage{geometry}
\usepackage{epstopdf}
\usepackage{hyperref}
\usepackage{color}
\usepackage{amsmath}

\journal{Journal of \LaTeX\ Templates}

\begin{document}

\begin{frontmatter}

\title{Time irreversibility and amplitude irreversibility measures for nonequilibrium processes}

\author[mymainaddress]{Wenpo Yao}
\author[myfirstaddress]{Jun Wang}
\author[myrdaddress,mythaddress]{Matja\v{z} Perc}
\author[mysecondaddress]{Wenli Yao}
\author[mythirdaddress]{Jiafei Dai}
\author[mymainaddress,myfaddress,myfiveaD]{Daqing Guo}
\author[mymainaddress,myfaddress,myfiveaD,mysixaD]{Dezhong Yao}

\address[mymainaddress]{The Clinical Hospital of Chengdu Brain Science Institute, MOE Key Lab for NeuroInformation, Center for Information in Medicine, School of Life Science and Technology, University of Electronic Science and Technology of China, Chengdu 611731, China}
\address[myfirstaddress]{Smart Health Big Data Analysis and Location Services Engineering Lab of Jiangsu Province, Nanjing University of Posts and Telecommunications, Nanjing 210023, Jiangsu, China}
\address[myrdaddress]{Faculty of Natural Sciences and Mathematics, University of Maribor, Koro\v{s}ka cesta 160, 2000 Maribor, Slovenia}
\address[mythaddress]{Complexity Science Hub Vienna, Josefst{\"a}dterstra{\ss}e 39, 1080 Vienna, Austria}
\address[mysecondaddress]{Department of hydraulic engineering, School of civil engineering, Tsinghua University, Beijing 100084, China}
\address[mythirdaddress]{Department of Neurology, Jinling Hospital, Medical School of Nanjing University, Nanjing 210002, Jiangsu, China}
\address[myfaddress]{Sichuan Institute for Brain Science and Brain-Inspired Intelligence, University of Electronic Science and Technology of China, Chengdu 611731, China}
\address[myfiveaD]{Research Unit of NeuroInformation, Chinese Academy of Medical Sciences,2019RU035, Chengdu, China}
\address[mysixaD]{School of Electrical Engineering, Zhengzhou University, Zhengzhou 450001, China}

\begin{abstract}
Time irreversibility, which characterizes nonequilibrium processes, can be measured based on the probabilistic differences between symmetric vectors. To simplify the quantification of time irreversibility, symmetric permutations instead of symmetric vectors have been employed in some studies. However, although effective in practical applications, this approach is conceptually incorrect. Time irreversibility should be measured based on the permutations of symmetric vectors rather than symmetric permutations, whereas symmetric permutations can instead be employed to determine the quantitative amplitude irreversibility -- a novel parameter proposed in this paper for nonequilibrium calculated by means of the probabilistic difference in amplitude fluctuations. Through theoretical and experimental analyses, we highlight the strong similarities and close associations between the time irreversibility and amplitude irreversibility measures. Our paper clarifies the connections of and the differences between the two types of permutation-based parameters for quantitative nonequilibrium, and by doing so, we bridge the concepts of amplitude irreversibility and time irreversibility and broaden the selection of quantitative tools for studying nonequilibrium processes in complex systems.
\end{abstract}


\end{frontmatter}

%\tableofcontents
\section{Introduction}
Complex systems, e.g., meteorological, economical or physiological systems, generally exhibit nonequilibrium processes, one of whose manifestations is a lack of time reversibility. Time reversibility is the property of invariance with respect to time reversal, and its lack is defined as time irreversibility (TIR) \cite{Weiss1975,Kelly1979}. TIR is widely used to detect nonlinearity, a necessary condition for chaotic behaviour; therefore, a quantitative TIR measure for complex processes is of particular interest.

Statistically speaking, to quantify TIR, one could measure the probabilistic differences either between forward and backward series or between symmetric vectors \cite{Ramsey1995,Yao2020ND}; the latter approach is preferable because of its real-time advantage. However, joint probability estimation for symmetric vectors is not trivial, particularly for unknown complex systems. Model-based probability estimators are generally linear ones that adopt the assumption of a particular distribution for the observed signals \cite{Xiong2017}, which is sometimes not appropriate for complex systems. Model-free probability estimators, such as kernel probability estimators, have been widely employed in informational analysis \cite{Xiong2017,Hlav2007,Marina2008}; however, they reconstruct time series, during which the symmetric or corresponding vectors cannot be matched for TIR quantification. Due to the limitations of traditional probability estimators, scholars have proposed some simplified methods for TIR quantification. Costa et al. \cite{Costa2005} quantified the TIR of heartbeats by measuring the difference between the average activation and relaxation energies, and they then simplified the parameter by distinguishing only the probabilistic difference between the up and down of the signal waveforms \cite{Costa2008}, a similar approach that has also been employed by Porta \cite{Porta2008}, Guzik \cite{Guzik2006} and Ehlers et al. \cite{Ehlers1998} for quantifying temporal asymmetry. Lacasa et al. \cite{Lacasa2012,Lacasa2015,Flanagan2016} quantified the TIR by measuring the divergence between the in- and out-degree distributions of either the original or the horizontal visibility graph. Additionally, symbolic approaches \cite{Daw2003}, such as approaches based on data compression \cite{Kennel2004}, false flipped symbols \cite{Daw2000}, and permutation \cite{Yao2020ND,Yao2019Ys,Yao2019E,Martin2019}, have received considerable attention for their simplicity, fast execution, noise insensitivity, etc. These simplified methods for TIR quantification play important roles in nonequilibrium analyses of complex systems.

Among these simplifications, approaches based on order patterns have been gaining popularity due to their close relation to multi-dimensional vectors and the inherent temporal structural information captured by the ordinal scheme \cite{Bandt2002,Bian2012,Yao2020}. However, in permutation-based measures, ignorance of particular issues might lead to failures in the quantification of the TIR. Due to the existence of forbidden permutations (i.e., permutations that cannot occur), some vectors or order patterns might not have corresponding symmetric forms \cite{Yao2020ND,Yao2019Ys,Yao2019E}, making the measurement of the probabilistic differences using division-based parameters (e.g., the relative entropy) unreliable. Equal values may lead to self-symmetric order patterns \cite{Yao2020ND,Yao2019E,Yao2020}; therefore, the equal states should not be broken by random perturbations or sorted according their order of occurrence in quantitative TIR. Some of \emph{the authors of the present paper} have incorrectly employed symmetric permutations (SPs) instead of symmetric vectors in TIR quantification, and they found that the SP-based methods have been shown to be effective for characterizing nonlinearity in model series and enabling the reliable detection of nonlinearity in real-world physiological\cite{Yao2019Ys}. However, it is conceptually incorrect to employ SPs in TIR quantification because the SP and permutations of symmetric vectors (PSVs) are not always the same \cite{Yao2020ND}. In other words, the simplified approach based on SPs quantifies other characteristics rather than the TIR.

In this contribution, we conduct comparative research on the relationship between SPs and PSVs, and on this basis, we define a novel statistical parameter, the amplitude irreversibility (AIR), for characterizing nonequilibrium from the perspective of amplitude fluctuations. This joint analysis of the AIR and TIR contributes to a better understanding of quantitative nonequilibrium and enhances current knowledge regarding the significance of equal values in permutation-based TIR and AIR analyses. Furthermore, a comprehensive discussion of the strong similarity between the AIR and TIR improves the understanding of quantitative nonequilibrium from a broader perspective.

\section{Materials \& methods}

\subsection{Quantitative time irreversibility and symmetric permutations}
Statistically speaking, a process $X(t)$ is said to be time reversible if $\{X(t_{1}),X(t_{2}),\cdots,X(t_{m})\}$ and $\{X(-t_{1}),X(-t_{2}),\cdots,X(-t_{m})\}$ have the same joint probability distribution for every $t_{1},t_{2},\cdots,t_{m}$ and $m$; otherwise, it is time irreversible \cite{Weiss1975}. In addition, $\{X(t_{1}),X(t_{2}),\cdots,X(t_{m})\}$ and $\{X(-t_{1}+n),X(-t_{2}+n),\cdots,X(-t_{m}+n)\}$ will also have the same joint probability distribution for every $n$ and $m$ if $X(t)$ is time reversible; in particular, if $n$=$t_{1}+t_{m}$, $\{X(t_{1}),X(t_{2}),\cdots,X(t_{m})\}$ have the same probability distribution as the symmetric process $\{X(t_{m}),\cdots,X(t_{2}),X(t_{1})\}$, i.e., the process will exhibit temporal symmetry \cite{Kelly1979,Ramsey1995}. Therefore, the probabilistic difference between the forward and backward processes and between the symmetric joint distributions of a process are equivalent for measuring the TIR \cite{Yao2020ND,Yao2019E}.

If the forward-backward method is employed, it is necessary to obtain and then reverse the whole process, which is not trivial and may even be impossible for a system that generates uninterrupted data. Therefore, methods based on symmetric vectors are preferable due to their advantage of real-time analysis \cite{Yao2020ND}.

As mentioned in the Introduction, the calculation of the probabilistic difference between symmetric vectors is not trivial, and the raw time series are generally simplified or reconstructed to quantify the TIR. Permutation-based methods are of particular interest because the ordinal scheme \cite{Bandt2002,Bian2012} does not require any modelling assumptions and is closely related to the TIR \cite{Yao2020ND,Yao2019Ys,Yao2019E}. In the ordinal scheme, the elements in the multi-dimensional vector $X_{m}^{\tau}(i)=\{ x(i),x(i+\tau),\ldots,x(i+(m-1)\tau)\}$, where $m$ is the dimension and $\tau$ is the delay, are reorganized (e.g., in ascending order, $x(j_1)<x(j_2)<\cdots<x(j_i)$), and the indexes of the reordered vector are used to formulate the permutation $\pi_{i}=\{j_{1},j_{2}, \cdots, j_{i}\}$.

As an alternative to raw symmetric vectors, SPs are directly employed by some of \emph{the present authors} for quantifying the TIR in some reports \cite{Yao2019Ys}, which is proved to be an effective approach for detecting nonlinearity in complex processes. However, we find that the PSVs are not always equivalent to the SPs and that methods based on SPs therefore do not correctly characterize the statistical concept of TIR.

\begin{figure}[htb]
  \centering
    \includegraphics[width=7cm,height=7cm]{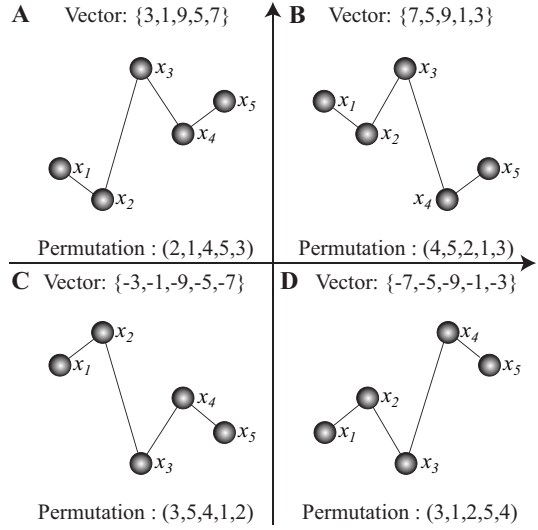}
  \caption{Relationships between symmetric vectors and their permutations. The vectors in \textbf{A} and \textbf{B} and those in \textbf{C} and \textbf{D} are symmetric with respect to the Y-axis; i.e., they exhibit temporal symmetry in terms of the X-axis angle. The vectors in \textbf{A} and \textbf{C} and those in \textbf{B} and \textbf{D} are symmetric with respect to the X-axis; i.e., they exhibit amplitude symmetry in terms of the Y-axis angle.}
  \label{fig1}
\end{figure}

Let us consider the difference between SPs and PSVs for an example case in which $m$=5, as illustrated in Fig.~\ref{fig1}.

Consider the vector \{3,1,9,5,7\} in Fig.~\ref{fig1}A; its permutation is (2,1,4,5,3), and its symmetric vector, depicted in Fig.~\ref{fig1}B, is \{7,5,9,1,3\}, whose order pattern is (4,5,2,1,3). However, (2,1,4,5,3) and (4,5,2,1,3) are not symmetric. The same is true for the vectors \{-3,-1,-9,-5,-7\} and \{-7,-5,-9,-1,-3\} shown in Fig.~\ref{fig1}C and~\ref{fig1}D; their permutations (i.e., (3,5,4,1,2) and (3,1,2,5,4)) are not symmetric either. For the example shown in Fig.~\ref{fig1}, to quantify the TIR, one should measure the probabilistic differences in the PSVs, i.e., the differences between (2,1,4,5,3) and (4,5,3,1,2) and between (3,5,4,1,2) and (3,1,2,5,4), rather than the probabilistic differences in the SPs, i.e., between (2,1,4,5,3) and (3,5,4,1,2) or between (4,5,2,1,3) and (3,1,2,4,5).

Therefore, it is incorrect to employ the probabilistic differences of SPs for quantifying TIR \cite{Yao2020ND}.

\subsection{The definition of amplitude reversibility}
Nevertheless, SPs have been reported to effectively characterize nonlinear activities in the related literature \cite{Yao2019Ys}. Therefore, we assume that there might be some associations between the PSVs, SPs and time reversibility. To determine the possible associations, let us dig further into the relationships among the vectors and permutations in Fig.~\ref{fig1}.

For the symmetric patterns (2,1,4,5,3) and (3,5,4,1,2), their corresponding vectors in Fig.~\ref{fig1}A and \ref{fig1}C are symmetric with respect to the X-axis; the same is true for the vectors corresponding to the symmetric patterns (4,5,3,1,2) and (3,1,2,5,4), as shown in Fig.~\ref{fig1}B and \ref{fig1}D. Whereas the vectors that characterize the TIR show Y-axis symmetry, say, time symmetry, the vectors that correspond to the SPs show X-axis symmetry, in other words, amplitude symmetry. Inspired by this finding, we define a novel statistical concept called amplitude reversibility, the quantification of which can be simplified by means of SPs.

\textbf{The definition of amplitude reversibility.} Given a process $X(t)$, we subtract its mean $\mu$ as $X(t)=X(t)-\mu$, if $\{X(t_{1}),X(t_{2}),\cdots,X(t_{m})\}$ and $\{-X(t_{1}),-X(t_{2}),\cdots,-X(t_{m})\}$ have the same joint probability distribution for every $t_{1},t_{2},\cdots,t_{m}$ and $m$, then $X(t)$ is amplitude reversible; otherwise, X(t) is amplitude irreversible. Thus, the AIR, which targets fluctuations in the amplitude of a dynamic process, is also a parameter that characterizes complex activities.

As in the case of quantifying the TIR, in some real-world situations, it is not feasible to obtain and take the negative of the whole process of interest. Therefore, to enable real-time processing, the probabilistic difference between $\{X(t_{1}),X(t_{2}),\cdots,X(t_{m})\}$ and $\{-X(t_{1}),-X(t_{2}),\cdots,-X(t_{m})\}$ could instead be simplified to a probabilistic difference between SPs.

AIR is a mathematical concept derived from the incorrect quantization of the TIR and from the relationship between the SPs and vectors, while it contains important physical significance, i.e., the nonequilibrium characteristic of amplitude fluctuations.

Fluctuation theorems provide some analytical expressions to describe nonequilibrium states \cite{Sevick2008}. In the Evans-Searles fluctuation theorem \cite{Evans1994,Evans2002,Searles2004}, the dissipation function $\Omega$ of observing trajectories is a dimensionless dissipated energy and is defined as arbitrary values $A$ and $-A$, their probabilistic difference $p(A)/p(-A)$ describes the asymmetry in the distribution of $\Omega$ over a particular ensemble of trajectories. According to the Evans-Searles fluctuation theorem, the vector and its amplitude-reverse form in the AIR could be defined as $A$ and $-A$, and the AIR describes amplitude fluctuations by the probabilistic difference $p(A)/p(-A)$. In the report of Costa et al. \cite{Costa2005}, they made assumptions that the transitions in systems require a specific amount of energy and quantified the nonequilibrium by the difference between activation and relaxation, which is then simplified by the number of increments and decrements \cite{Costa2008}, a special case of permutation AIR when $m$=2. The order patterns, as a reliable alternative to the original vector, naturally arise from the time series and inherit structural information of vectors, and the permutation AIR is a reliable parameter to characterize the asymmetry in the distribution of permutations over an ensemble of ordinal trajectories.

\subsection{Equal-value permutation and vectors}
For the permutation TIR, equal values may lead to self-symmetric order patterns with important statistical implications, i.e., time reversibility or temporal symmetry \cite{Yao2020ND,Yao2019E,Yao2020}. Hence, equal values might also be expected to have similar implications in the case of the AIR measure.

In the original ordinal scheme and some variants thereof, equal values can be eliminated by adding small random perturbations or can be arranged in the order of occurrence; however, these approaches might yield false findings or misleading conclusions if there are a large number of equal values \cite{Yao2019E,Yao2020,Yao2019P}. Bian et al. \cite{Bian2012} proposed a desirable alternative for considering the contributions of equal states. In this improved permutation scheme, equal values are first placed in adjacent positions their order of occurrence, i.e., $\cdots<x(j_k)=x(j_l)< \cdots <x(j_x)=x(j_y)=x(j_z)<\cdots$, and the indexes of all the members of each such group of equal values in the original permutation, i.e., $\pi_{i}=\{\cdots,j_k,j_l,\cdots,j_x,j_y,j_z,\cdots\}$, are then set to be equal to the lowest index in the group, i.e., $\pi_{i}=\{\cdots,j_k,j_k,\cdots,j_x,j_x,j_x,\cdots\}$.

\begin{figure}[htb]
  \centering
    \includegraphics[width=7cm,height=7cm]{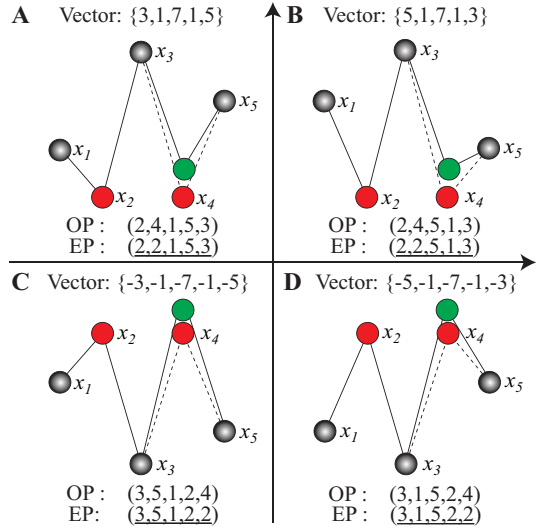}
  \caption{Example of the effects of equal values on symmetric vectors and their permutations. The vectors in \textbf{A} and \textbf{B} and those in \textbf{C} and \textbf{D} are temporally symmetric, and the vectors in \textbf{A} and \textbf{C} and those in \textbf{B} and \textbf{D} are amplitude symmetric. The values in red are equal, and those shown in green are the alternatives if we organize the equal values according to the order of occurrence. The original (denoted by 'OP') and equal-value (underlined and denoted by 'EP') permutations are shown below each subplot.}
  \label{fig2}
\end{figure}

The example illustrated in Fig.~\ref{fig2} suggests that equal-value permutations play a crucial role in the formulation of the order patterns for the quantitative AIR measure.

For the vector depicted in Fig.~\ref{fig2}A, i.e., \{3,1,7,1,5\}, if we organize the two equal values (the '1's) according to the order of occurrence, then the permutation is (2,4,1,5,3), which is also the order pattern of the vector \{3,1,7,2,5\}. Therefore, the original order pattern does not reliably reflect the relationship among the elements of the vector, which is shared among the other three subplots. More importantly, for the vectors symmetric about the X-axis in Fig.~\ref{fig2}A and in Fig.~\ref{fig2}C, i.e., \{3,1,7,1,5\} and \{-3,-1,-7,-1,-5\}, respectively, their original permutations (2,4,1,5,3) and (3,5,1,2,4) are not symmetric. If we instead employ the equal-value ordinal scheme, the order patterns of the vectors in Fig.~\ref{fig2}A and ~\ref{fig2}C are (2,2,1,5,3) and (3,5,1,2,2), respectively, which are symmetric; similarly, the order patterns of the vectors in Fig.~\ref{fig2}B and~\ref{fig2}D (i.e., (2,2,5,1,3) and (3,1,5,2,2), respectively) are also symmetric. Therefore, equal-value permutations are crucial for the simplified quantitative AIR measure.

The equal-value ordinal scheme more reliably represents the temporal structure of a dynamic process and more accurately reflects the relationship among the elements in a vector than the original ordinal scheme. Because the only unique feature of equal-value permutation is that the indexes of equal values are rewritten, if there are no equal states, the equal-value ordinal method reduces to the original one.

Overall, the equal-value ordinal scheme is particularly relevant for the determination of the permutation TIR and AIR and should be the preferred choice in permutation analyses.

\subsection{Permutation TIR and AIR}
\begin{figure}[htb]
  \centering
    \includegraphics[width=8.5cm,height=7cm]{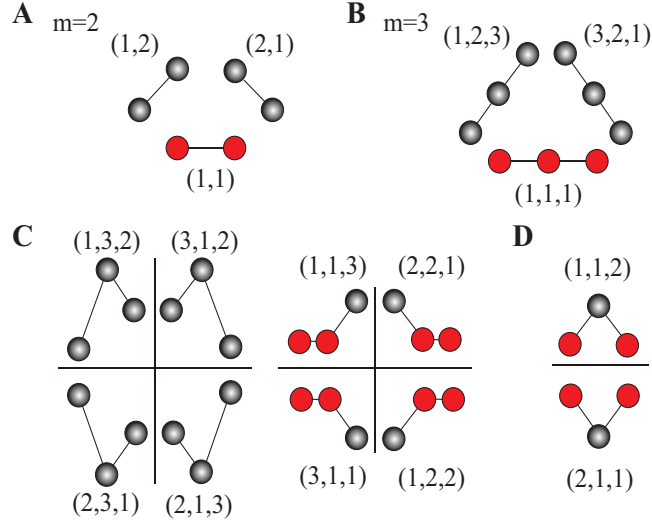}
  \caption{Vectors and their permutations when $m$=2 and 3. \textbf{A} and \textbf{B} display centrally symmetric vectors when $m$=2 and $m$=3, and \textbf{C} shows two groups of time- and amplitude-reverse vectors. The pair of amplitude-reverse vectors in \textbf{D} are both temporal self-symmetric. Red elements indicate equal values in vectors.}
  \label{fig3}
\end{figure}

Let us further analyze the difference between AIR and TIR, and elucidate the physical meaning of AIR. Basically speaking, the TIR quantifies the probabilistic difference from the time reversal or temporal symmetry perspective, while the AIR measures the probabilistic divergence in amplitude fluctuations. As for permutation TIR and AIR, the relationship of vectors and their order patterns when $m$=2 and 3 are illustrated in Fig.~\ref{fig3}.

The PSVs and the SPs are generally different; however, in some special cases, i.e., in the case of centrally symmetric vectors, they are the same. When $m$=2 in Fig.~\ref{fig3}A, the three order patterns, namely, up (1,2), down (2,1) and equality (1,1), are all centrally symmetric, which is true for the case of $m$=3 in Fig.~\ref{fig3}B and other dimensions. And with the increase of $m$, there will be more pairs of centrally symmetric vectors. Note that the whole-equality vector, whose permutation is (1,1,1,$\cdots$), implies both time and amplitude reversibility.

The difference between the AIR and TIR lies in the Fig.~\ref{fig3}C and Fig.~\ref{fig3}D where vectors are not centrally symmetric. We need to calculate the probabilistic differences of time-reverse vectors for TIR while calculate those of amplitude-reverse vectors for TIR. The patterns (1,3,2) and (2,3,1) form a pair of SPs whose vectors are amplitude reversed, while the PSV of (1,3,2) is (3,1,2) and that of (2,3,1) is (2,1,3), and their vectors are time reversed or temporally symmetry. As for the pair of special vectors (1,1,2) and (2,1,1) in Fig.~\ref{fig3}D, they are temporal self-symmetric that their symmetric vectors are themselves, and they have particular physical implication, i.e., time reversibility; however, quantitative AIR requires to calculate they probabilistic difference. The different pair of vectors targeted by the AIR and TIR is their crucial difference.

Theoretically speaking, the larger length of vectors, the more amount of permutations, and the larger difference between the permutation TIR and AIR. In real-world applications, the difference between the TIR and AIR and the extent of the difference are affected by the probability distributions of permutations, which will be analyzed in following sections.

For the permutation TIR and AIR, the existence of forbidden permutations makes employing subtraction-based parameters, e.g., the parameter Ys defined in Eq.~\ref{eq1}, more reliable for calculating the probabilistic differences between paired order patterns. In the expression for Ys \cite{Yao2020ND,Yao2019Ys,Yao2019E,Yao2020}, $p(\pi_{i})$ and $p(\pi_{j})$ denote the probabilities of the SPs or PSVs, respectively, and $p(\pi_{i})$ should not be smaller than $p(\pi_{j})$.

\begin{eqnarray}
\label{eq1}
Ys \langle p(\pi_{i}),p(\pi_{j}) \rangle= p(\pi_{i})\frac{p(\pi_{i})-p(\pi_{j})}{p(\pi_{i})+p(\pi_{j})}
\end{eqnarray}

\section{Results}
In this section, model series and their surrogate data are generated to test the permutation TIR and AIR, which are then applied to quantify the nonequilibrium characteristics of two kinds of real-world physiological signals.

\subsection{TIR and AIR in model series}
\begin{figure}[htb]
  \centering
    \includegraphics[width=14cm,height=5cm]{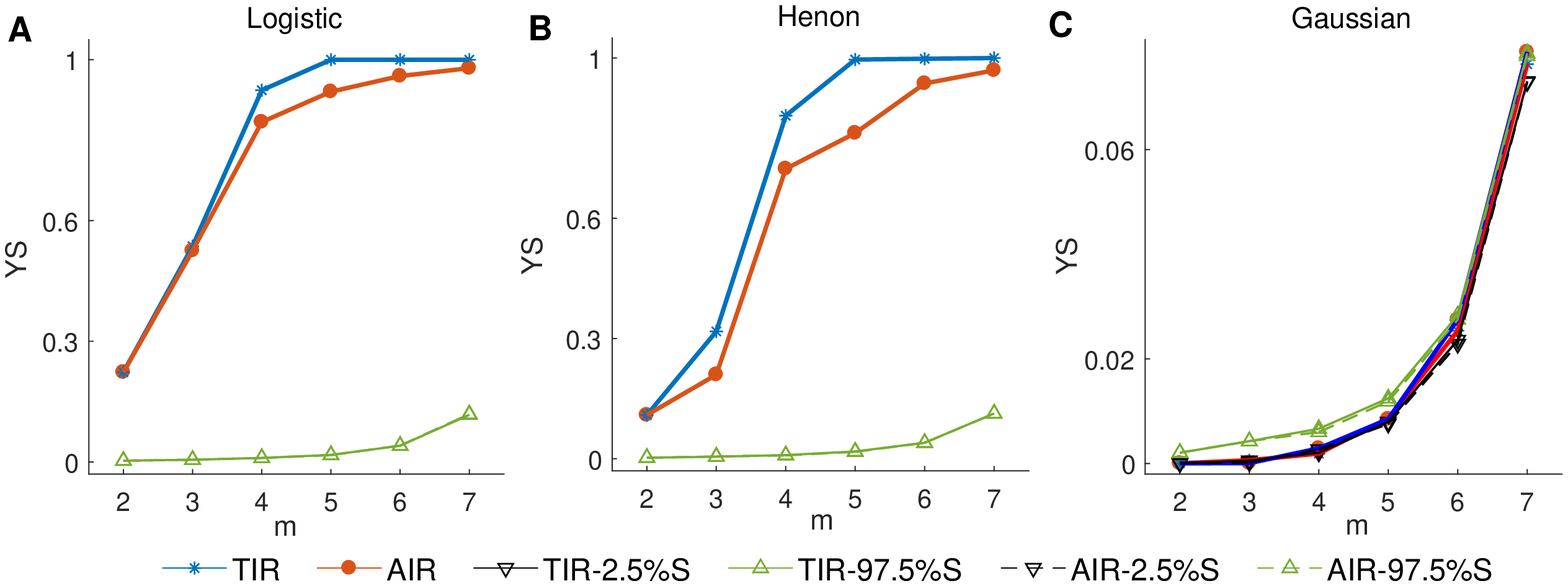}
  \caption{TIR and AIR measures for logistic, Henon and Gaussian series. For the chaotic logistic and Henon series, $x_{1}$=$y_{1}$=0.01, $r$=4, $\alpha$=1.4, $\beta$=0.3, and x-components with a data length of 20$\times$(7!)=100800 are applied.}
  \label{fig4}
\end{figure}

The logistic, Henon and Gaussian model series are employed to test the permutation TIR and AIR, and the results are illustrated in Fig.~\ref{fig4}. The logistic equation, $x_{t+1}=r \cdot x_{t} (1- x_{t})$, and the coupled Henon equations, $x_{t+1}=1-\alpha \cdot x^{2}_{t}+y_{t}$ and $y_{t+1}=\beta \cdot x_{t}$, were used to generate nonlinear chaotic series, and zero-mean Gaussian series were employed as linear series. We also generated 500 sets of linear surrogate data for each set of model data using the improved amplitude-adjusted Fourier transform algorithm \cite{Schreiber1996} to test the nonlinearity detection abilities of the TIR and AIR.

As shown in Fig.~\ref{fig4}, the TIR and AIR values for the logistic and Henon series are all larger than the 97.5th percentile of the surrogate data, whereas those for the Gaussian series are between the 2.5th and 97.5th percentiles of the surrogate data, suggesting that both the TIR and AIR enable effective nonlinearity detection according to surrogate theory \cite{Yao2020ND,Yao2019Ys}.

The permutation TIR and AIR have consistent results while different trends in the three model series. When $m$=2, permutation TIR and AIR are the same for each model series, and when $m$ increases from 3 to 5, TIR and AIR values show different increase trends. For the two chaotic series, the probabilistic differences between the PSVs are larger that those between SPs that TIR is larger than the AIR. Moreover, due to the influence of forbidden permutations \cite{Yao2020ND,Yao2019Ys}, when $m$ is greater than or equal to 5, the differences between permutation AIR and TIR decrease. The permutation TIR is equal to 1, indicating that no order pattern of a corresponding symmetric vector exists, while the AIR is nonzero when $m$=7, suggesting that there are still SPs although very rare. The different trends of permutation TIR and AIR in chaotic series suggest the AIR contains different information about non-equilibrium properties of the system from TIR, and the extent of their difference is determined by the probability distributions of permutations.

The permutation AIR and TIR share strong similarities as well as differences in model series, and they are both effective for the detection of nonlinearity. We should note that the TIR measure, i.e., the probabilistic differences in the temporal reversible process, and the AIR measure, i.e., the probabilistic differences in the amplitude fluctuations, extract different types of characteristics of complex systems, although they obtain similar results.

\subsection{TIR and AIR in physiological time series}
In previous studies, an AIR measure based on the original permutation scheme has been applied to brain signals \cite{Yao2019Ys}, and a TIR measure based on the equal-value permutation scheme has been used to characterize heartbeat signals \cite{Yao2019E,Yao2020}. In this paper, we conduct a comparative analysis of the permutation TIR and AIR for these physiological time series.

To collect electroencephalograms (EEGs), 22 epileptic patients (aged 15 to 49 yrs. old, mean 26.95$\pm$8.91 yrs. old) were recruited from Jinling Hospital (JLH); all the patients had been seizure-free for approximately 20 to 30 days. In addition, 22 healthy people (aged 4 to 51 yrs. old, mean 30.0$\pm$13.1 yrs. old) were also enrolled for EEG collection. Following the standard 10-20 EEG system, 16 scalp electrodes were placed on the subjects in their idle states; the duration of data collection was approximately 1 minute, and the sampling frequency was 512 Hz. Briefly, these epileptic data were from our previous study, and detailed information can be found in \cite{Yao2019Ys}.

For the heartbeat signals, data were obtained from the public PhysioNet database \cite{Goldb2000}. Of these heartbeat data, 44 sets were derived from electrocardiograms (ECGs) recorded from patients with congestive heart failure (CHF) (aged 22 to 79, mean 55.5$\pm$11.4 yrs. old), 20 sets were obtained from healthy elderly people (aged 68 to 85, mean 74.5$\pm$4.4 yrs. old), and 20 sets were obtained from healthy young people (aged 21 to 34, mean 25.8$\pm$4.3 yrs. old).

\begin{figure}[htb]
  \centering
    \includegraphics[width=15cm,height=7cm]{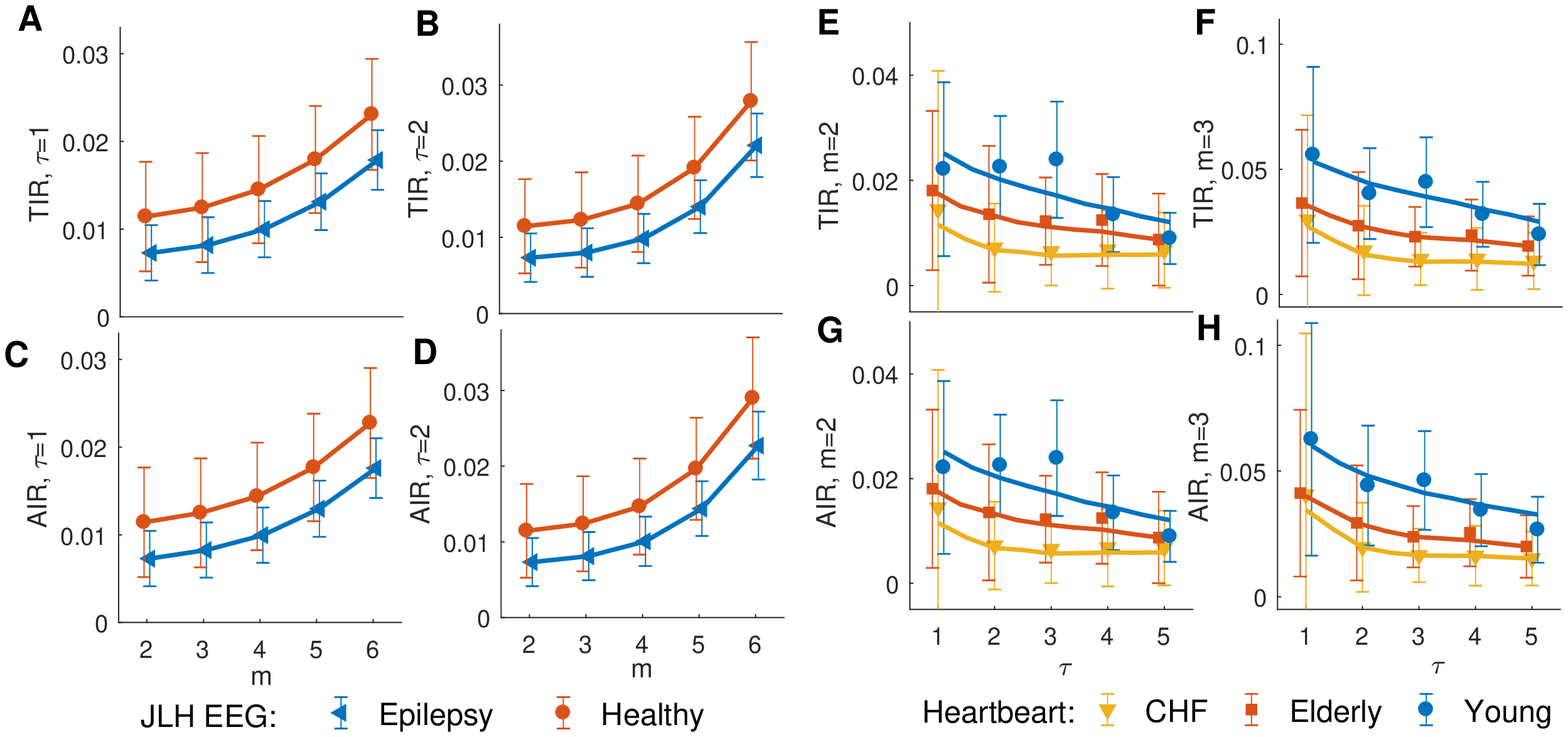}
   \caption{AIR and TIR results based on equal-value permutations for two sets of EEGs and three sets of heartbeats. \textbf{A} and \textbf{B} show the TIR results for the EEGs of the patients with epilepsy and healthy people for $\tau$=1 and 2, respectively, and $m$ from 2 to 6; \textbf{C} and \textbf{D} present the corresponding AIR results. \textbf{E} and \textbf{F} show the TIR results for the heartbeats of the patients with CHF, the healthy elderly individuals and healthy young individuals when $m$=2 and 3, respectively, and $\tau$ ranging from 1 to 5; \textbf{G} and \textbf{H} present the corresponding AIR results.}
  \label{fig5}
\end{figure}

The AIR and TIR values for the EEGs and heartbeats are displayed in Fig.~\ref{fig5}.

Similar to the results for the model series, the permutation AIR and TIR show some differences but still enable highly similar nonequilibrium detections in both EEGs and heartbeats.

For the two groups of EEGs, the permutation AIR and TIR share consistent outcomes: the seizure-free EEGs of the patients with epilepsy show lower nonequilibrium than those of their healthy counterparts (p$<$0.005). Moreover, the equal-value AIR results for the EEGs of the patients with epilepsy from JLH are similar to and yield the same conclusion as those for the AIR measure based on the original permutation scheme \cite{Yao2019Ys}. For example, when $m$=3 and $\tau$=2, the AIR values of the EEGs of the healthy people based on the equal-value and original-order patterns are 0.0123 and 0.0122, respectively, whereas those of the EEGs of the patients with epilepsy are 0.0080 and 0.0072, respectively. The reason for this similarity lies in the fact that equal values occur very rarely in neuro-electrical signals, accounting for only 0.36\% and 0.05\% of the values in the EEGs of the patients with epilepsy and the healthy controls, respectively, and the distributions of these equal values therefore have no significant effects on the results. Nevertheless, the equal-value ordinal scheme is recommended due to its physical importance for both AIR and TIR \cite{Yao2020ND,Yao2020}.

Epilepsy is a life-threatening neurological disorder characterized by recurrent seizures. The characteristic seizures manifest as the sudden development of abnormal, excessive, and synchronous neuronal firing. Abnormally large nonlinearities in seizure EEGs have been widely reported \cite{Yao2020ND}; however, the lower AIR and TIR values observed in our work physiologically suggest long-term negative effects on brain activity, as the neural disorder causes declinations in the nonequilibrium of brain electrical activity \cite{Yao2019Ys,Yao2014}.

For the three groups of heartbeat signals, complexity-loss theory can be applied to both the TIR and AIR measures based on equal-value permutation. The central postulate of complexity-loss theory \cite{Costa2005,Costa2008,Yao2019E,Yao2020,Goldb2002CL,Ivanov1999} is that healthy physiological systems exhibit a type of nonlinear complexity that degrades with age and disease along with the accompanying reduction in adaptive capability. According to this theory, healthy young heartbeats should exhibit the largest nonlinearity, while CHF heartbeats should exhibit smallest nonlinearity, and the complexity of healthy elderly heartbeats should lie in between. The results are consistent with these expectations, and the difference between each pair of heartbeat groups is statistically significant (p$<$0.001), particularly when $\tau$$>$1 \cite{Yao2019E,Yao2020}.

The permutation AIR and TIR in the two kinds of physiological signals, although obtaining highly similar results, are different in the manner they detect nonequilibrium characteristics, sharing the findings in the analysis of model series. The highly consistent findings regarding the TIR and AIR, i.e., temporal asymmetry and amplitude irreversibility, indicate that the joint analysis of different nonequilibrium aspects might further broaden the quantitative nonequilibrium, which will be discussed in the following section.

According to the test results for the model series and physiological data, the simplified AIR and TIR measures both effectively characterize the nonequilibrium property of complex systems. These two closely related parameters, although they yield highly consistent findings, are different in the manner they quantify nonequilibrium: the TIR measure is a traditional parameter that targets the time-reversible probabilistic differences, while the AIR measure involves nonequilibrium amplitude fluctuations.

\section{Discussion}
In this paper, we analysed the relationship between permutations and vectors and conducted a comparative analysis of the TIR and AIR measures. From our tests, we find that the similarities and differences between the permutation AIR and TIR and their physical significance in quantitative nonequilibrium require further discussion.

\begin{figure}[htb]
  \centering
    \includegraphics[width=15cm,height=6cm]{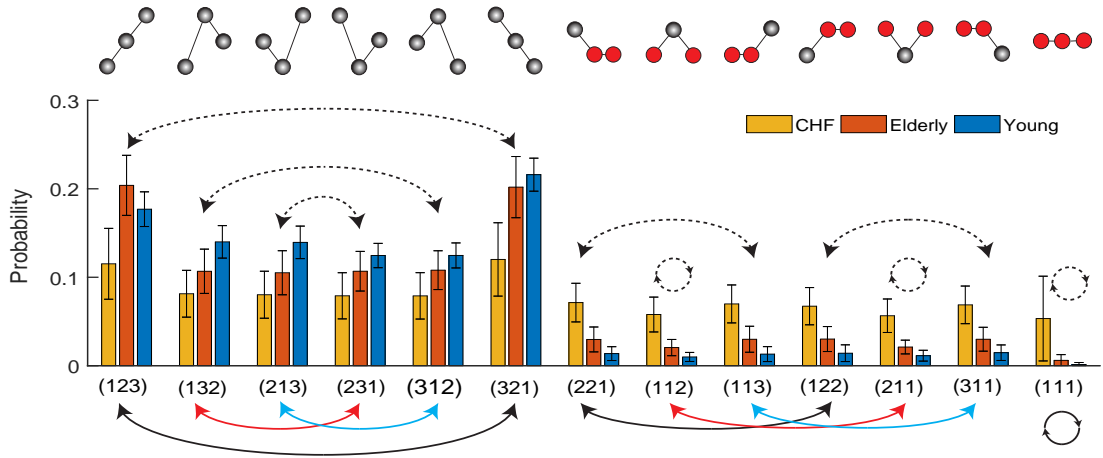}
    \caption{Order patterns ($m$=3, $\tau$=3) of the heartbeats of the patients with CHF, the healthy elderly individuals and the healthy young individuals. The vectors of their corresponding permutations above the figure are in line with the Fig.~\ref{fig3}, and equal values in the vectors are indicated in red. The dashed double arrows above the plot point to pairs of PSVs for TIR quantification, and the solid double arrows below the plot point to pairs of SPs for AIR quantification. The three ordered patterns above the figure and the order pattern below it represented by circular arrows correspond to time reversibility and amplitude reversibility, respectively.}
  \label{fig7}
\end{figure}

Let us further identify the connections between SPs and PSVs and between the AIR and TIR measures based on the distributions of the order patterns. As mentioned in section 2.4, when $m$=2, there are only three relationships, namely, up, down and equality, and they are all centrally symmetric, i.e., the SPs and PSVs, as well as the simplified quantitative AIR and TIR measures, are the same, which has been demonstrated by the test results. When $m$=3, we illustrate the exemplary probability distributions of the order patterns ($\tau$=3) of the three groups of heartbeats in Fig.~\ref{fig7}. The relationships of vectors and permutations further confirm the difference between and the connections of the PSVs and SPs in Fig.~\ref{fig3}. The SPs and PSVs only share the pair comprising (123) and (321) as well as (111) whose vectors are centrally symmetric; however the permutation TIR and AIR still have high consistent results in heartbeats. The TIR values for the heartbeats of the patients with CHF, the healthy elderly individuals and healthy young individuals are 0.0141, 0.0230 and 0.0448, respectively, and the corresponding AIR values are 0.0164, 0.0238 and 0.0462, respectively, which are very close to the TIR values. The TIR measures the probabilistic differences in time-reversible permutations and the AIR calculates the probabilistic differences in amplitude-reversible order patterns; i.e., the two parameters target different aspects of nonequilibrium, suggested by the differences between the paired SPs and PSVs and between the AIR and TIR. In addition, the similarity between the probabilistic differences between SPs and PSVs contributes to the similarity in the AIR and TIR measures. For different dimensions and delays in the ordinal scheme, the SPs and PSVs show slight differences and high similarity; therefore, the different aspects of nonequilibrium contribute to the difference between the AIR and TIR measures, while the consistency between the PSVs and SPs for the model series, EEGs and heartbeats is the direct reason for the high similarity in the permutation TIR and AIR.

However, although the similarities between the SPs and PSVs are consistent with those between the AIR and TIR measures, they are not the root cause of the latter. TIR measure quantifies the probabilistic differences from the time-reversible perspective, and the AIR measure quantifies the differences in amplitude fluctuations, and they both target the same characteristic, i.e., nonequilibrium, which is the fundamental reason for these highly consistent results. This similarity inspires us to focus on the nonequilibrium itself rather than particular measures thereof, i.e., the TIR or AIR measures. In the Introduction, we note that the traditional kernel estimators are not suitable for TIR due to the reconstruction of probability distributions; however, according to our findings, we can still detect probabilistic differences during time series transformations, such as between the positive and negative parts in the Heaviside kernel function \cite{Xiong2017}. Moreover, visibility graphs, although they have been gaining popularity in the field of TIR quantification \cite{Lacasa2012,Lacasa2015,Flanagan2016,Donges2013}, also reconstruct the time series. From this perspective, the difference between the in- and out-degree distributions of the visibility graph is also not directly consistent with the statistical concept of TIR, and it in fact measures a kind of networked nonequilibrium. According to the visibility graph TIR, we can derive other kinds of either local or global networked nonequilibrium by constructing weighted and directional networks and by calculating the probabilistic difference between the in and out degrees of nodes. These approaches all effectively characterize nonlinear dynamic processes because they detect nonequilibrium rather than TIR. Based on this fact, we should broaden our view of the measurement of nonequilibrium rather than limiting ourselves to the concept of time reversibility or amplitude reversibility.

Fluctuation theorems contribute to our understanding of how irreversibility emerges from reversible dynamics \cite{Sevick2008}, and both the TIR and AIR are effective parameters to characterize the fluctuations in nonequilibrium processes. The differences between and similarities of the permutation TIR and AIR contribute to improving the understanding and quantification of the nonequilibrium characteristics from a broader perspective and to more widely exploration to the essential characteristics of the nonequilibrium nature of complex systems.

\section{Conclusions}
To conclude, we have corrected the conceptual error underlying the use of SPs in the quantification of TIR and have defined a novel concept of amplitude reversibility for quantitative nonequilibrium analogous to time reversibility, and we have also demonstrated the significance of equal-value permutation for both TIR and AIR. The AIR and TIR target different aspects of nonequilibrium fluctuations; however, the permutation TIR and AIR show high consistency on both model series and real-world data; inspired by this observation, we clarify the relationship of AIR and TIR and quantitative nonequilibrium.

Our findings have more profound significance that we should not allow ourselves to be limited by the statistical definition of the TIR and that we can instead directly target the nonequilibrium of a system by considering the differences in various time series conversions, thus broadening the scope of nonequilibrium analysis.

\section{Acknowledgment}
The project is supported by the National Natural Science Foundation of China (Grant Nos. 61527815, 31771149, 81571770, 61933003), the CAMS Innovation Fund for Medical Sciences CIFMS (Grant No. 2019-I2M-5-039), Sichuan Science and Technology Program (Grant No. 2018HH0003), China Postdoctoral Science Foundation (2020M683279), and by the Slovenian Research Agency (Grant Nos. J4-9302, J1-9112, and P1-0403).

\section{Conflict of interest}

The authors declare that they have no conflict of interest.

\nocite{*}

\bibliography{mybibfile}% Produces the bibliography via BibTeX.

\end{document}